\documentstyle[12pt,subeqn]{article}

\newcommand\vek[1]{\mbox{\rmfamily\bfseries\itshape#1}}
\newcommand\vekexp[1]{\mbox{\scriptsize\rmfamily\bfseries\itshape#1}}
\def\rd{{\rm d}}

\begin{document}

\title{Surface Tension of a Metal-Electrolyte Boundary: 
Exactly Solvable Model}

\author{
L. {\v S}amaj$^{1,2}$ and B. Jancovici$^1$ 
}

\maketitle

\begin{abstract}
An ideal conductor electrode in contact with a semi-infinite 
two-dimensional two-component plasma in an external potential 
is considered.
The model is mapped onto an integrable sine-Gordon theory
with Dirichlet boundary conditions.
The information gained from the mapping provides an explicit 
form of the surface tension in the plasma-stability regime.  
\end{abstract}


\vskip 0.7truecm
\noindent LPT Orsay 00-80 

\vfill

\noindent $^1$ Laboratoire de Physique Th{\'e}orique, Universit{\'e}
de Paris-Sud, B{\^a}timent 210, 91405 Orsay Cedex, France (Unit{\'e}
Mixte de Recherche no. 8627 - CNRS);

\noindent e-mail: Bernard.Jancovici@th.u-psud.fr and 
samaj@th.u-psud.fr

\noindent $^2$ On leave from the Institute of Physics, Slovak Academy
of Sciences, Bratislava, Slovakia; 

\noindent e-mail: fyzimaes@savba.sk

\newpage

\section{Introduction}
In a previous paper \cite{Samaj}, hereafter referred to as I, the bulk
thermodynamic properties (free energy, specific heat, etc...) of a
model, the two-dimensional (2D) two-component plasma (TCP), 
or Coulomb gas, have been obtained, exactly. 
In the present paper, a surface property of the same model is
considered: the surface tension, at a rectilinear 
interface between an ideal conductor and the Coulomb gas, is obtained,
exactly, as a function of the bulk density, the applied electric
potential, and the temperature, in the whole temperature range for 
which the point-particle model is stable. 
Like in I, a mapping onto a sine-Gordon field theory, now with a 
Dirichlet boundary condition, is made, and known results about 
that field theory are used. The resulting surface tension is checked 
on its high-temperature expansion derived from a renormalized Mayer 
expansion and on its singular behaviour close to the collapse point.
 
The model under consideration mimics the interface between an
electrolyte (the two-component plasma, made of two species of
point-particles, of opposite charges $\pm 1$) and an electrode (the
ideal conductor). 
Classical equilibrium statistical mechanics is used.     
In the grand-canonical formalism, the control parameters are the
inverse temperature $\beta$ and the two fugacities $z_+$ and $z_-$ 
of the positive and negative particles, respectively. 
Instead of $z_+$ and $z_-$, it is convenient to use $z$ and $\varphi$ 
defined by $z_{\pm}= ze^{\pm\beta\varphi}$. 
Alternatively, chemical potentials $\mu_+$ and $\mu_-$ can be defined 
by $z_{\pm}=\exp(\beta\mu_{\pm})/\lambda^2$, where $\lambda$ is the 
de Broglie thermal wavelength. 
The bulk properties depend only \cite{Lieb} on the chemical
potential combination $\mu=(\mu_++\mu_-)/2$, i.e. on $z$,
while $\mu_+-\mu_-$ (or $\varphi$) is relevant only for the surface
properties \cite{Jancovici}, \cite{Cornu}, \cite{Forrester}. 
The parameter $\varphi$ has a physical meaning: it is the 
electric-potential difference between the bulk and the electrode. 
Indeed, if the potential of the electrode is taken as the
zero and $\varphi$ is the potential in the bulk, each chemical
potential, i.e the reversible work for adding a positive or negative
particle into the bulk, has an electric part $\varphi$ or $-\varphi$,
respectively.
    
The point-particle model is stable against collapse of positive-negative
pairs for $\beta<2$. 
This is also the stability range in presence of a
rectilinear ideal conductor wall. 
Indeed, a particle at a distance $x$ 
from the wall interacts with its own image through a potential 
$(1/2)\ln(2x)$ and the corresponding Boltzmann factor $(2x)^{-\beta/2}$ 
is integrable at small $x$ if and only if $\beta<2$ (at large $x$, the
interaction is screened).

The paper is organized as follows. 
In Section 2, the model is precisely defined and its mapping 
onto a sine-Gordon field theory is made.
This field theory is described in Section 3. 
The desired surface tension is derived in Section 4. 
Its high temperature expansion is checked in Section 5.
Its singular behaviour close to the collapse point $\beta = 2$
is checked in Section 6.

\section{Mapping}
We consider an infinite 2D space of points ${\vek r} \in R^2$ defined
by Cartesian coordinates $(x,y)$.
The model electrode-electrolyte interface is localized along the
$y$ axis, namely at $\{ {\vek r} = (0,y) \}$.
The half-space $x<0$ is assumed to be occupied by an ideal conductor
of dielectric constant $\epsilon \to \infty$, impenetrable to particles.
The electrolyte in the complementary half-space $x>0$ is modeled by
the classical 2D TCP of point particles $\{ j\}$ of charge 
$\{ q_j = \pm 1\}$, immersed in a homogeneous medium 
of dielectric constant $=1$.
The interface is kept at zero potential while one assumes 
a given potential $\varphi$ in the bulk.
Equivalently, there is a splitting of the fugacities of 
the $\pm$ particles:
\begin{equation} \label{1}
z_+ = z \exp(\beta \varphi) , \quad  
z_- = z \exp(-\beta \varphi) 
\end{equation}
The system is translationally invariant in the $y$ direction, 
so the position-dependent particle densities $n_{\pm}({\vek r})$
depend only on $x$.
Let us denote their asymptotical $x\to\infty$ values by $n_+ = n_-
= n/2$ where $n$ is the total particle number density.
In the case $\varphi=0$, $n_+(x) = n_-(x)$ everywhere.

In infinite space, the Coulomb potential $v$ at spatial position 
${\vek r}$, induced by a unit charge at the origin, 
is given by the 2D Poisson equation
\begin{equation} \label{2}
\Delta v({\vek r}) = - 2 \pi \delta({\vek r})
\end{equation}
The solution of (\ref{2}) reads
\begin{equation} \label{3}
v({\vek r}) = - \ln (\vert {\vek r} \vert / r_0)
\end{equation}
where the length constant $r_0$, which fixes the zero point of energy,
is set for simplicity to unity.
Here, the interaction energy $E$ of particles 
$\{ q_j, {\vek r}_j=(x_j>0,y_j)\}$ consists of two parts
(see, e.g., \cite{Jackson}):

\begin{subequations}
(i) direct particle-particle interactions, 
\begin{equation} \label{4a}
\sum_{i<j} q_i q_j v(\vert {\vek r}_i - {\vek r}_j \vert)
\end{equation}

(ii) interactions of particles with the images of other particles and
with their self-images due to the presence of the conducting wall,
\begin{equation} \label{4b}
- {1\over 2} \sum_{i,j} q_i q_j v( \vert {\vek r}_i - {\vek r}_j^*
\vert)
\end{equation}
\end{subequations}
where ${\vek r}^* = (-x,y)$.
Introducing the microscopic charge density $\rho({\vek r}) =
\sum_j q_j \delta({\vek r} - {\vek r}_j)$, the energy contributions
(\ref{4a}) and (\ref{4b}) are expressible as follows
\begin{equation} \label{5}
E = {1\over 2} \int \rd^2 r \int \rd^2 r' \rho({\vek r})
\left[ v({\vek r},{\vek r}') - v({\vek r}^*,{\vek r}') 
\right] \rho({\vek r}') - {1\over 2} N v(0) 
\end{equation}
where $v(0)$ is the self-energy.
Introducing the microscopic charge + image charge density
\begin{equation} \label{6}
{\bar \rho}({\vek r}) = \sum_j q_j \left[ \delta(x-x_j) - \delta(x+x_j)
\right] \delta(y-y_j)
\end{equation}
the energy (\ref{5}) can be rewritten into a more convenient form
\begin{equation} \label{7}
E = {1\over 4} \int \rd^2 r \int \rd^2 r' {\bar \rho}({\vek r})
v({\vek r},{\vek r}') {\bar \rho}({\vek r}') - {1\over 2} N v(0) 
\end{equation}  
where the integrations over ${\vek r}$ and ${\vek r}'$ are now
taken over the whole space.

The thermodynamic characteristics of the system are determined 
by the grand partition function $\Xi$,
\begin{subequations}
\begin{equation} \label{8a}
\Xi = \sum_{N_+ =0}^{\infty} \sum_{N_- =0}^{\infty}
{z_+^{N_+}\over N_+!} {z_-^{N_-}\over N_-!} Q(N_+,N_-) 
\end{equation}
with
\begin{equation} \label{8b}
Q(N_+,N_-) = \int \prod_{j=1}^N \rd^2 r_j 
\exp \left[ -\beta E(\{ q_j,{\vek r}_j\})\right]
\end{equation}
\end{subequations}
being the canonical partition function of $N_+$ positive and
$N_-$ negative charges and $N = N_+ + N_-$.
To express $\Xi$ in terms of a 2D Euclidean sine-Gordon theory,
we first recall that $-\Delta/(2\pi)$ is the inverse operator of
$v({\vek r})$ [see eq. (\ref{2})].
The standard identity then follows
\begin{equation} \label{9}
\exp \left[ - {\beta\over 4} \int \rd^2 r \int \rd^2 r'
{\bar \rho}({\vek r}) v({\vek r},{\vek r}') {\bar \rho}({\vek r}')
\right] = {\int {\cal D} \phi \exp \left[ \int \rd^2 r \left(
{1\over 2} \phi \Delta \phi + {\rm i} \sqrt{\pi\beta}\phi
{\bar \rho} \right) \right] \over \int {\cal D} \phi
\exp \left( \int \rd^2 r {1\over 2} \phi \Delta \phi \right)}
\end{equation} 
where $\phi({\vek r})$ is a real scalar field and $\int {\cal D}
\phi$ denotes the functional integration over this field.
Inserting ${\bar \rho}$ from (\ref{6}), the second term in
the action of the field theory (\ref{9}) takes a nonlocal form
${\rm i} \sqrt{\pi\beta} \sum_j q_j [\phi(x_j,y_j) - \phi(-x_j,y_j)]$.
It is therefore convenient to reformulate the field theory (\ref{9})
as a boundary problem using a procedure proposed in ref. \cite{Callan}.
One introduces two new fields
\begin{subequations} \label{10}
\begin{eqnarray} 
\phi_e(x,y) & = & {1\over \sqrt{2}} \left[ \phi(x,y) + \phi(-x,y)
\right] \label{10a} \\
\phi_o(x,y) & = & {1\over \sqrt{2}} \left[ \phi(x,y) - \phi(-x,y)
\right] \label{10b}
\end{eqnarray}
\end{subequations}
defined only in the positive $x\ge 0$ half-space.
Clearly, the even field has Neumann boundary conditions 
$\partial \phi_e(x,y)/\partial x \vert_{x=0} = 0$ and
the odd field has Dirichlet boundary conditions
$\phi_o(x=0,y) = 0$.
It is straightforward to show that
\begin{equation} \label{11}
\int \rd^2 r {1\over 2} \phi \Delta \phi = {1\over 2} 
\int_{x>0} \rd^2 r \left( \phi_e \Delta \phi_e +
\phi_o \Delta \phi_o \right)
\end{equation}
The even field contributes to the action only by its free-field
part $\phi_e\Delta \phi_e/2$, which is ``cancelled'' with its
counterpart in the denominator of (\ref{9}).
Thus, when $\phi_o$ is renamed as $\phi$, the rhs of (\ref{9}) 
is expressible as a local field theory 
formulated in the half-space $x>0$:
\begin{equation} \label{12}
{\int {\cal D} \phi \exp \left\{ \int_{x>0} \rd^2 r
\left[ - {1\over 2} (\nabla \phi)^2 + {\rm i} \sqrt{2\pi \beta}
\sum_j q_j \phi({\vek r}_j) \right] \right\} \over
\int {\cal D} \phi \exp \left[ - \int_{x>0} \rd^2 r {1\over 2}
(\nabla\phi)^2 \right]}
\end{equation} 
with Dirichlet boundary conditions $\phi(x=0,y)=0$.
Now, defining by $\langle \cdots \rangle$ the average over the
field theory (\ref{12}), one proceeds along the standard line,
i.e., express $\Xi$ as follows
\begin{equation} \label{13}
\Xi = \sum_{N_+ =0}^{\infty} \sum_{N_- =0}^{\infty}
{{\bar z}^{N_+ +N_-} {\rm e}^{\beta \varphi (N_+ - N_-)} \over 
N_+! N_-!} \Big\langle 
\left( \int_{x>0} \rd^2 r ~ {\rm e}^{{\rm i} \sqrt{2\pi\beta} 
\phi({\vekexp r})} \right)^{N_+}
\left( \int_{x>0} \rd^2 r ~ {\rm e}^{-{\rm i} \sqrt{2\pi\beta} 
\phi({\vekexp r})} \right)^{N_-} \Big\rangle
\end{equation}
where ${\bar z} = z \exp[\beta v(0)/2]$ is the
fugacity renormalized by the self-energy term, and 
afterwards sum over $N_+$ and $N_-$.
The final result reads
\begin{equation} \label{14}
\Xi = {\int {\cal D} \phi \exp \left\{ \int_{x>0} \rd^2 r
\left[ - {1\over 2} (\nabla \phi)^2 + 2 {\bar z} \cos 
(\sqrt{2\pi\beta}\phi - {\rm i}\beta\varphi) \right] \right\} \over
\int {\cal D} \phi \exp \left[ - \int_{x>0} \rd^2 r {1\over 2}
(\nabla\phi)^2 \right]}
\end{equation}
with the fixed value of the field at the boundary, 
$\phi(x=0,y) = 0$.

The field theory in (\ref{14}) can be viewed as the ordinary
2D Euclidean sine-Gordon model in the half-space,
defined by the action
\begin{subequations} \label{15}
\begin{equation} \label{15a}
{\cal A}_{sG} = \int_{x>0} \rd^2 r \left[ - {1\over 2} (\nabla \phi)^2
+ 2 {\bar z} \cos (\beta_{sG} \phi) \right]
\end{equation}
with 
\begin{equation} \label{15b}
\beta_{sG} = \sqrt{2\pi\beta} 
\end{equation}
and Dirichlet boundary conditions $\phi(x=0,y)=\phi_0$, 
\begin{equation} \label{15c}
\phi_0 = - {\rm i} \sqrt{{\beta\over 2\pi}} ~ \varphi
\end{equation}
\end{subequations}   
The boundary value $\phi_0$ usually enters into the formalism 
in the combination
\begin{equation} \label{16}
\xi = {4\pi \over \beta_{sG}} \phi_0 = - 2 {\rm i} \varphi
\end{equation}
which will also be used in what follows.
For $\varphi$ real (which is the case of interest), we assume
that the solution of the present theory corresponds to an analytical 
continuation of the results of the theory with real Dirichlet
boundary conditions.

Without going into details we mention that one can proceed as
above also in the case of the 2D TCP in contact with
a dielectric of dielectric constant $=0$. In that case,
the images have the same charges as the source particles, 
the even field (\ref{10a}) becomes the relevant one, and that
gives a mapping onto an equivalent boundary sine-Gordon 
model with Neumann boundary conditions.  
 
\section{Boundary sine-Gordon model}
The integrability property of the bulk (infinite in both $x,y$-directions)
sine-Gordon model along the standard lines of the Bethe ansatz technique
is well known from the seventies \cite{Zamolodchikov1}.
The discrete symmetry of the theory $\phi \to \phi + 2\pi n /\beta_{sG}$
($n$ integer) is spontaneously broken in the domain
$0< \beta_{sG}^2 < 8\pi$; one has to consider one of infinitely many
ground states $\{ \vert 0_n\rangle\}$ characterized by
$\langle \phi \rangle_n = 2\pi n/\beta_{sG}$, say the one with $n=0$.
The spectrum of particles involves solitons $S$, antisolitons ${\bar S}$
and soliton-antisoliton bound states (breathers)
$\{ B_j, j=1,2,\ldots <1/q \}$ with masses
\begin{equation} \label{17}
m_j = 2 M \sin \left( {j q \pi \over 2} \right)
\end{equation}
where $M$ is the soliton mass. 
The parameter $q$, defined by
\begin{equation} \label{18}
q = {\beta_{sG}^2 \over 8\pi - \beta_{sG}^2}
\end{equation}
ranges from $0$ to $\infty$: the breathers exist for $q\in(0,1)$
(or in the plasma stability range $0<\beta<2$) and there are 
no breathers for $q$ larger than 1 ($2<\beta<4$).

In the seminal work \cite{Ghoshal} it was argued that the restriction
of the sine-Gordon model to the half space $(x>0,y)$ does not break
its integrability if one adds a boundary action term
\begin{equation} \label{19}
({\cal A}_{sG})_B = \int_{-\infty}^{\infty} \rd y ~ m 
\cos \left[ {\beta_{sG} \over 2} (\phi_B - \phi_0) \right]
\end{equation}
where $m$ and $\phi_0$ are free parameters, and
$\phi_B(y) = \phi(x,y)\vert_{x=0}$.
The underlying sine-Gordon theory (\ref{15}) corresponds to
the $m\to\infty$ limit of (\ref{19}) which sets $\phi_B$ to $\phi_0$.

The free energy of the theory (\ref{15}) is related to the
ground state energy of the boundary quantum $(1+1)$-dimensional
sine-Gordon model, defined by Lagrangian
\begin{equation} \label{20}
{\cal L}_{sG} = \int_0^L \rd x \left[ {1\over 2} (\partial \phi)^2
+ 2 {\bar z} \cos (\beta_{sG}\phi) \right] 
\end{equation}
$L\to \infty$, with boundary conditions $\phi(0) = \phi_0$ and 
$\phi(L) = \phi_0'$ considered in terms of $\xi$ and $\xi'$, 
respectively [see eq. (\ref{16})].
There exists a lattice regularization of the theory (\ref{20}),
namely the XXZ model in boundary magnetic fields 
\cite{Alcaraz}, defined by Hamiltonian
\begin{equation} \label{21}
{\cal H}_{XXZ} = \epsilon {\tau \over 2\pi \sin \tau} 
\left\{ \sum_{j=1}^{N-1}
\left[ \sigma_j^x \sigma_{j+1}^x + \sigma_j^y \sigma_{j+1}^y -
(\cos \tau) \sigma_j^z \sigma_{j+1}^z \right] + h \sigma_1^z
+ h' \sigma_N^z \right\}
\end{equation}
with $\tau\in (0,\pi/2)$. 
Here, $\epsilon = 1$ and $\epsilon = -1$ correspond to the
ferromagnetic and antiferromagnetic cases of the XXZ-chain,
respectively.
This model results as the hamiltonian limit
of the inhomogeneous 6-vertex model on an open strip 
\cite{Destri}, with an alternating imaginary part 
$\pm {\rm i}\Lambda$ added to the spectral parameter on
alternating vertices.
The continuum scaling limit is given by taking $\Lambda \to \infty$,
$N\to \infty$, and the lattice spacing $a\to 0$, such that
$L \equiv N a$ remains finite.
In the bulk, the regularization fixes
$\sqrt{{\bar z}} \propto (1/a) \exp(-{\rm const} \times \Lambda)$
and
\begin{subequations} \label{22}
\begin{eqnarray}  
\beta_{sG}^2 & = & 8 \tau \quad \quad (\epsilon =1) \label{22a} \\
\beta_{sG}^2 & = & 8 (\pi - \tau) \quad \quad (\epsilon = - 1) 
\label{22b}
\end{eqnarray}
\end{subequations}
As $\tau\in (0,\pi/2)$, the ferromagnetic regime corresponds
to $q\in (0,1)$ and the antiferromagnetic regime to $q\in (1,\infty)$.
As concerns the interrelation between the ``surface'' quantities
\cite{Fendley}, one defines the function
\begin{subequations}
\begin{equation} \label{23a}
f(a,b) = - {\rm i} \ln \left( {\sinh (({\rm i}b-a)/2) \over
\sinh (({\rm i}b+a)/2)} \right)
\end{equation} 
and 
\begin{equation} \label{23b}
H \equiv {1\over \tau} f({\rm i}\tau,-{\rm i}\ln(h+\cos\tau)) 
\end{equation}
\end{subequations}
Then, introducing an auxiliary variable
\begin{equation} \label{24}
t = {\pi \over \tau} \quad \quad  t\ge 2
\end{equation}
one has
\begin{subequations} \label{25}
\begin{eqnarray} 
\xi & = & {\pi \over 2} \left( t -1 - H \right)
\quad \quad (\epsilon = 1)  \label{25a} \\
\xi & = & {\pi \over 2} \left( 1 - {H \over t-1}
\right) \quad \quad (\epsilon = -1) \label{25b}
\end{eqnarray}
\end{subequations}
The same formula hold for $\xi'$ in terms of $H'$, resp. $h'$.

To study boundary effects in the XXZ model (\ref{21}),
one looks for the solutions of the Bethe equations which
correspond to a wave-function localized at $j=0$ or $j=N$
and exponentially decreasing away from the boundary.
These boundary bound states were identified with new boundary
strings in the Bethe ansatz \cite{Skorik1}, \cite{LeClair}
(for a review see \cite{Skorik2}).
In the thermodynamic limit $N\to\infty$, when the left
and the right boundaries can be treated independently
and the overlap of the corresponding wave-functions is
negligibly small, the lattice ground state energy was found
as a function of the parameter $\Lambda$ of the related 
6-vertex model.
In the continuum limit $\Lambda \to \infty$ corresponding to 
the $(1+1)$-dimensional sine-Gordon model (\ref{20}),
the ground state energy of the last was obtained in the form
\begin{equation} \label{26}
E_{ground} = E_{bulk} + E_{bdries} + O(1/L)
\end{equation} 
where
\begin{subequations} \label{27}
\begin{eqnarray}
E_{bulk} & = & - L {M^2 \over 4} \tan \left( {\pi \over 2(t-1)}
\right)  \label{27a} \\
E_{bdries} & = & - {M\over 2} \left[ 
{\sin \left( H\pi/[2(t-1)]\right) \over \sin \left( t\pi/[2(t-1)]
\right)} +
{\sin \left( H'\pi/[2(t-1)]\right) \over \sin \left( t\pi/[2(t-1)]
\right)} \right. \nonumber \\
& & \left. \quad - \cot \left({t\pi \over 4 (t-1)} \right) - 1 \right]
\label{27b}
\end{eqnarray}
\end{subequations}
for the ferromagnetic case $\epsilon =1$ \cite{Skorik1} (note a
different notation in this reference) and
\begin{subequations} \label{28}
\begin{eqnarray}
E_{bulk} & = & L {M^2 \over 4} \cot \left( {t\pi \over 2}
\right)  \label{28a} \\
E_{bdries} & = & - {M\over 2} \left[ 
{\sin \left( (t- H)\pi/2 \right) \over \sin \left( t\pi/2 \right)} +
{\sin \left( (t- H')\pi/2 \right) \over \sin \left( t\pi/2 \right)} -
\cot \left({t\pi \over 4 } \right) - 1 \right]
\label{28b}
\end{eqnarray}
\end{subequations}
for the antiferromagnetic case $\epsilon = - 1$ \cite{LeClair}.
In terms of the sine-Gordon parameters, ``bulk'' $q$ (\ref{18}) and 
``boundary'' $\xi, \xi'$ (\ref{25}), both Eqs. (\ref{27}) and 
Eqs. (\ref{28}) take the same form
\begin{subequations} \label{29}
\begin{eqnarray}
E_{bulk} & = & - L {M^2 \over 4} \tan \left( {q \pi \over 2}
\right)  \label{29a} \\
E_{bdries} & = & - {M\over 2} \left\{ 
{1\over \cos(q\pi/2)} \left[ \cos(q\xi) + \cos(q\xi') - 1 \right]
+ \tan \left( {q \pi \over 2} \right) - 1 \right\}
\label{29b} 
\end{eqnarray}
\end{subequations}

\section{Surface tension of the plasma}
For a Coulomb gas of volume $V$ bounded by a surface of area $S$,
the grand potential $\Omega = - \beta^{-1} \ln \Xi$ is the sum
of a volume part and a surface part:
\begin{equation} \label{30}
\Omega = - V p(z,\beta) + S \gamma(z,\beta)
\end{equation}
where $p$ is the pressure and $\gamma$ the surface tension.
For a strip $L \times R$, $R\to \infty$ and $L$ large,
the ``specific'' $\Omega/R$ is given by
\begin{equation} \label{31}
\lim_{R\to\infty} {\Omega \over R} = - L p(z,\beta) +
\gamma(z,\beta)\vert_{x=0} + \gamma(z,\beta)\vert_{x=L}
+O(1/L) 
\end{equation}
The thermodynamics of (\ref{14}) is mapped onto the ground state 
of (\ref{20}) according to
\begin{equation} \label{32}
\beta \Omega = R E_{ground}
\end{equation}
In the considered $L\to\infty$ limit, the boundary energy (\ref{29b}) 
is the sum of two clearly separated contributions comming from 
the boundaries at $x=0$ and $x=L$. 
To calculate the surface tension, one keeps only 
the contribution at $x=0$, and identifies $\xi$ with the
bulk potential $\varphi$ via eq. (\ref{16}). 
Thus, one gets
\begin{subequations} \label{33}
\begin{eqnarray} 
\beta p & = & {M^2 \over 4} \tan \left( {q\pi \over 2} \right) \\
\beta \gamma & = &  - {M\over 4} \left\{ 
{1\over \cos(q\pi/2)} \left[ 2 \cos(q\xi) - 1 \right]
+ \tan \left( {q \pi \over 2} \right) - 1 \right\}
\end{eqnarray}
\end{subequations}

Inserting (\ref{15b}) into (\ref{18}) and considering (\ref{16}),
the sine-Gordon parameters are expressible in terms of the Coulomb
ones as follows
\begin{equation} \label{34}
q = {\beta \over 4-\beta}, \quad \quad \xi = - 2 {\rm i} \varphi
\end{equation}
As concerns the link between the soliton mass 
$M$ and the fugacity $z$, the formalism of section 2 showed us 
that $z$ renormalizes multiplicatively.
To give $z$ a precise meaning, one has to fix the normalization
of the field $\cos(\beta_{sG}\phi)$. 
The conformal normalization proposed in refs. \cite{Zamolodchikov2}
and \cite{Lukyanov} corresponds to the short-distance limit
of the two-point correlation function
\begin{equation} \label{35}
\langle \cos(\beta_{sG}\phi)({\vek x})
\cos(\beta_{sG}\phi)({\vek y}) \rangle \to {1\over 2} \vert 
{\vek x} - {\vek y} \vert^{-\beta_{sG}^2/(2\pi)}
\end{equation}
This normalization, equivalent to a well known leading 
short-distance behaviour of the positive-negative pair correlation 
in the Coulomb gas, fixes the $z$-$M$ relationship as follows
\begin{equation} \label{36}
z = {\Gamma(q/(q+1)) \over \pi \Gamma(1/(q+1))} \left[
M {\sqrt{\pi} \Gamma((q+1)/2) \over 2 \Gamma(q/2)} 
\right]^{2/(q+1)}
\end{equation}
The total particle number density, generated via
\begin{equation} \label{37}
n = z {\partial \beta p \over \partial z}
\end{equation}
is related to $M$ as follows
\begin{equation} \label{38}
n = {1\over 4} M^2 (1+q) \tan \left( {\pi q \over 2} \right)
\end{equation}
The $n$-$z$ relationship is given in paper I, Eqs. (49), (50).
The singular behaviour of $n$ as $\beta\to 2$ ($z$ fixed) 
can be deduced from these formulae:
\begin{equation} \label{39}
n \sim {4 z^2 \pi \over 2-\beta}
\end{equation}

Finally, we have
\begin{equation} \label{40}
\beta p = n \left( 1 - {\beta \over 4} \right)
\end{equation}
and
\begin{eqnarray} \label{41}
\beta \gamma &  =  & - {1\over 2} \left[ {n(4-\beta) \over
2 \sin(\pi\beta/(4-\beta)) } \right]^{1/2}
\left\{  2 \cosh \left( {2\beta \varphi \over 4-\beta} \right)
-1 \right. \nonumber \\
& & \left. \hskip5cm +
\sin \left( {\pi \beta \over 2 (4-\beta)} \right) -
\cos \left( {\pi \beta \over 2 (4-\beta)} \right) \right\}
\end{eqnarray}
$\beta \gamma$ has the small $\beta$-expansion
\begin{equation} \label{42}
\beta \gamma = - {1\over 8} (2\pi\beta n)^{1/2} \left\{
1 + \left[ {\pi \over 16} + {2 \varphi^2 \over \pi} \right] \beta
+ \left[ {\pi \over 64} \left( 1+{\pi\over 6} \right) + {\varphi^2
\over 2\pi} \right] \beta^2 + \ldots \right\}
\end{equation}
With regard to (\ref{39}), for $z$ fixed, $p$ and $\gamma$ exhibit
the same type of collapse singularities as $\beta \to 2^-$:
\begin{equation} \label{43}
p \sim {z^2 \pi \over 2-\beta}
\end{equation}
and
\begin{equation} \label{44}
\gamma \sim - {z \cosh (2\varphi) \over 2-\beta}
\end{equation}

\section{High-temperature expansion}
As a check of the exact expression (\ref{41}) of the surface tension, 
the beginning of its high-temperature expansion [expansion in powers of 
$\beta$, eq. (\ref{42})]  will now be compared to a direct evaluation 
of the two first terms of this expansion derived from the renormalized 
Mayer expansion of the free energy of the Coulomb system. 
In the following, all functions and integrals are defined in 
the half-space $x\geq 0$. 

The surface tension $\gamma$ can be defined as the boundary part per
unit length of the grand potential $\Omega$. 
The total numbers of positive and negative particles, respectively, 
are $N_+=-\beta z_+\partial\Omega/\partial z_+$ and
$N_-=-\beta z_-\partial\Omega/\partial z_-$. 
Going to the variables $z$ and $\varphi$ defined by 
$z_\pm =ze^{\pm \beta \varphi}$ gives for the total number of particles
$N=N_++N_-=-\beta z\partial \Omega/\partial z$. 
The boundary part of this relation is 
\begin{equation} \label{45}
-\beta z \frac{\partial \gamma}{\partial z}
=\int_0^{\infty} \rd x\,[n(x)-n] 
\end{equation}
Since, as recalled in I, $z$ is proportional to $n^{1-\beta/4}$, 
(\ref{45}) can be rewritten as 
\begin{equation} \label{46}
-\beta n \frac{\partial \gamma}{\partial n}= \left( 1-\frac{\beta}{4}
\right) \int_0^{\infty}\rd x\,[n(x)-n] 
\end{equation}
This relation (\ref{46}) will be used for computing the surface tension
$\gamma$ from the density profile $n(x)$ which will be determined as a 
function of the bulk density $n$, the electric potential $\varphi$ in 
the bulk, and the inverse temperature $\beta$.

Our starting point is the relation obeyed by the density profiles:
\begin{equation} \label{47}
\ln \left[\frac{n_{\pm}(1)}{z_{\pm}(1)}\right]=
\frac{\delta \Delta [n]}{\delta n_{\pm}(1)} 
\end{equation}
where $\Delta [n]$ is the negative of the excess free energy times 
$\beta$, considered as a functional of the position-dependent densities,
and $z_{\pm}(1)$ is the position-dependent fugacity at point 1.
Since a particle at a distance $x$ from the boundary has an interaction
$(1/2)\ln(2x)$ with its own image,
\begin{equation} \label{48}
z_{\pm}(x)=z\exp[\pm\beta\varphi-(\beta/2)\ln(2x)] 
\end{equation}
The relation (\ref{47}) is exact. Here, the renormalized Mayer
expansion, described in I, is used for expanding $\Delta [n]$ up to
order $\beta^2$, i.e., we keep only the contributions 
${\bar D}_0$ (two field circles connected by a simple $-\beta v$
bond plus the sum of ring diagrams) and the graph $D_1$. 
However the latter one will be shown to give no contribution to 
the density profiles at the desired order, 
and we shall be left with only ${\bar D}_0$, which constitutes 
an approximation of the Debye-H\"uckel type.
However, here, the interaction $v(1,2)$ between particles 1 and 2
includes the contributions from the images, i.e., 
$v(1,2) = - \ln r_{12} + \ln r_{12}^*$ where
$r_{12}$ is the distance between points 1 and 2, $r_{12}^*$ is
the distance between point 1 and the image of point 2.
The contribution of ${\bar D}_0$ to the functional derivative in
(\ref{47}) is
\begin{equation} \label{49}
\frac{\delta {\bar D}_0 [n]}{\delta n_{\pm}(1)}=\mp\beta\int \rd 2\, 
v(1,2) [n_+(2)-n_-(2)]+\frac{1}{2}[K(1,1)+\beta v(1,1)] 
\end{equation}
where $K(1,2)$ is the renormalized bond defined by the integral
equation 
\begin{equation} \label{50}
K(1,2)=-\beta v(1,2)+\int \rd 3 [-\beta v(1,3)]n(3)K(3,2) 
\end{equation}
This integral equation (\ref{50}) can be transformed into a partial
differential equation by taking the Laplacian with respect to 1,
$\Delta_1$, of both sides of (\ref{50}), and using 
$\Delta_1 v(1,2)= -2\pi \delta(1,2)$, which gives
\begin{equation} \label{51}
\Delta_1 K(1,2)=-2\pi\beta\delta(1,2)+2\pi\beta n(1)K(1,2)
\end{equation}
Since $v(1,2)$, the Coulomb interaction in presence of a
conducting wall, vanishes when 1 is on the wall, the same boundary
condition holds for $K(1,2)$.

The above equations can be solved for the density profile $n(x)$ 
by iterations, starting with the lowest-order approximation of a 
constant $n(x)=n$ in (\ref{51}). 
Then, the solution of (\ref{51}), with its boundary condition, is
\begin{equation} \label{52}
K^{(0)}(1,2)= -\beta K_0(\kappa r_{12})+\beta K_0(\kappa r_{12}^*) 
\end{equation}
where $\kappa^2=2\pi\beta n$ ($\kappa$ is the inverse Debye length), 
and $K_0$ is the modified Bessel function of second kind. 
Using this lowest-order $K^{(0)}$ in (\ref{49}), approximating 
$\Delta[n]$ by ${\bar D}_0[n]$ and using (\ref{48}) in (\ref{47}), 
gives
\begin{equation} \label{53}
n_{\pm}(x)=z\exp \left\{ \pm\beta[\varphi-\varphi(x)]
+\frac{1}{2}\lim_{r_{12}\rightarrow 0}[-\beta K_0(\kappa r_{12})
-\beta\ln r_{12}]+\frac{1}{2}\beta K_0(2\kappa x) \right\}
\end{equation}
where, with point 1 at a distance $x$ from the boundary,
\begin{equation} \label{54}
\varphi(x)=\int \rd 2 \, v(1,2)[n_+(2)-n_-(2)]
\end{equation}
$\varphi(x)$ is the electric potential created at point 1 by the
charge distribution (localized near the boundary) $n_+(2)-n_-(2)$.
$\varphi-\varphi(x)$ is a finite quantity which goes to zero as 
$x\rightarrow\infty$. Since, at the Debye-H\"uckel order of
approximation, the bulk fugacity and density are related by
\begin{equation} \label{55}
n=2z\exp\{(1/2)\lim_{r_{12}\rightarrow 0}[-\beta K_0(\kappa r_{12})
-\beta\ln r_{12}]\} 
\end{equation}
(\ref{53}) can be rewritten as
\begin{equation} \label{56}
n_{\pm}(x)=\frac{n}{2}\exp\{\pm\beta[\varphi-\varphi(x)]+
(\beta/2) K_0(2\kappa x)\}
\end{equation}
(\ref{56}) has a simple physical interpretation: each particle feels a
mean one-body potential made of two parts: $\varphi(x)$ is the
electric potential created by the surface charge 
density in the plasma, $-(1/2)K_0(2\kappa x)$ is the screened
interaction of the particle with its image. 
The linearized form of (\ref{56}) gives for the density profile
\begin{equation} \label{57}
n(x)=n_+(x)+n_-(x)=n \left[ 1+\frac{1}{2}\beta K_0(2\kappa x) \right]
\end{equation}
In the following, the integrals
\begin{subequations} \label{58}
\begin{equation} \label{58a}
\int_0^{\infty}K_0(x) \rd x =\frac{\pi}{2} 
\end{equation}
and
\begin{equation} \label{58b}
\int_0^{\infty}K_0^2(x) \rd x =\frac{\pi^2}{4}
\end{equation}
\end{subequations}
will be needed. 
Using (\ref{57}) and (\ref{58a}) in (\ref{46}) gives the surface
tension $\gamma$ at lowest order in $\beta$: 
\begin{equation} \label{59}
\beta\gamma =-\frac{1}{8}(2\pi\beta n)^{1/2}
\end{equation}
At this order in $\beta$, an explicit form of $\varphi(x)$ was not 
needed for computing the surface tension, neither does that surface tension
depend on the parameter $\varphi$. 
However, $\varphi(x)$ will be needed in the following. 
It can be easily obtained by writing, from the
linearized form of (\ref{56}), $n_+(x)-n_-(x)=\beta n[\varphi-\varphi(x)]$ 
and using the Poisson equation $d^2[\varphi-\varphi(x)]/dx^2=
2\pi[n_+(x)-n_-(x)]$ with the boundary condition $\varphi(0)=0$. 
One obtains
\begin{equation} \label{60}
\varphi-\varphi(x)=\varphi\exp(-\kappa x)
\end{equation}

The next iteration for $n(x)$ is obtained by using the density 
(\ref{56}) in the equation (\ref{51}) for $K$ and treating 
$n(x)-n=(1/2)n\beta K_0(2\kappa x)$ as a perturbation. 
Now $K=K^{(0)}+K^{(1)}$, where $K^{(0)}$ is defined by (\ref{52}). 
To first order in the density perturbation, (\ref{51}) and (\ref{52}) give
\begin{equation} \label{61}
(\Delta_1-\kappa^2)K^{(1)}(1,2)=2\pi\beta\delta n(1)K^{(0)}(1,2)
\end{equation}
where $\delta n(1)=(1/2)n\beta K_0(2\kappa x)$ with $x$ the
distance of point 1 to the boundary. 
The solution of (\ref{61}), with the boundary condition that
$K^{(1)}(1,2)$ vanishes when 1 is on the wall, is studied, by the method 
of Green functions, in the Appendix, where it is shown 
that $K^{(1)}(1,1)$ is a function of the coordinate $x_1$ of 1 such that
\begin{equation} \label{62}
\int_0^{\infty} \rd x_1K^{(1)}(1,1)=
\frac{\beta^2}{16\kappa} \left( \pi-\frac{\pi^2}{4} \right)
\end{equation}
It is a priori necessary to keep also the contribution from the graph 
$D_1$ to $\Delta[n]$. 
However, at the present order in $\beta$, $\delta D_1[n]/\delta 
n_{\pm}(1)$ can be evaluated for constant densities and then it
vanishes. 
Taking also $K^{(1)}$ into account, we now have, 
instead of (\ref{56}),
\begin{equation} \label{63}
n_{\pm}(x)=\frac{n}{2}\exp \left\{ \pm\beta[\varphi-\varphi(x)]+
\frac{1}{2} [\beta K_0(2\kappa x)+K^{(1)}(x)] \right\}
\end{equation}
where $K^{(1)}(1,1)$, when 1 has the coordinate $x$, is renamed 
$K^{(1)}(x)$.
Expanding the exponential in (\ref{63}) to order $\beta^2$ gives for 
the total density $n_++n_-$ 
\begin{equation} \label{64}
n(x)-n=\frac{n\beta}{2}K_0(2\kappa x)+\frac{n}{2}K^{(1)}(x)+
\frac{n\beta^2}{8}K_0^2(2\kappa x)+
\frac{n\beta^2}{2}[\varphi-\varphi(x)]^2 
\end{equation}
where it is sufficient to use for $\varphi-\varphi(x)$ the
lower-order expression (\ref{60}). 
Thus, using (\ref{58}) and (\ref{62}), one finds
\begin{equation} \label{65}
\int_0^{\infty}\rd x\,[n(x)-n]=\frac{\kappa}{16} \left[ 
1+\frac{\beta}{4}\left( 1- \frac{\pi}{4} \right) +
\frac{\beta\pi}{8} + \frac{2\beta}{\pi}\varphi^2 \right]
\end{equation}
Using (\ref{65}) in (\ref{46}) gives the final result
\begin{equation} \label{66}
\beta\gamma=-\frac{1}{8}(2\pi\beta n)^{1/2} \left[ 1+
\left( \frac{\pi}{16}+\frac{2\varphi^2}{\pi}\right) 
\beta+O(\beta^2) \right]
\end{equation}
in agreement with (\ref{42}).

\section{Collapse singularity}
To check the singular behaviour of the surface tension $\gamma$ 
near the collapse point $\beta=2$, eq. (\ref{44}), 
one starts with the exact $x\to 0$ limits
\begin{equation} \label{67}
n_+(x) - n_+ \sim {z_+ \over (2x)^{\beta/2}} , \quad \quad
n_-(x) - n_- \sim {z_- \over (2x)^{\beta/2}} 
\end{equation}
which can be derived directly by using the grand canonical
or canonical formalisms, in analogy with the short-distance 
expansion of the positive-negative pair correlation in the bulk.
At $\beta = 2$, the exact result for the density profile
reads \cite{Cornu}
\begin{equation} \label{68}
n_{\pm}(x) - n_{\pm} = {m\over 2\pi} \int_0^{\infty}
\rd l \left[ - {m\over \kappa_l} + {\kappa_l \exp(\pm \beta 
\varphi) + m \over m \cosh(\beta \varphi) + \kappa_l} \right]
\exp (-2\kappa_l x)
\end{equation}
where $m = 2\pi z$ and $\kappa_l =(m^2+l^2)^{1/2}$ (note that 
in the original work \cite{Cornu} there are some mistakes in
the equation (3.25)).
The $x\to 0$ limit of eq. (\ref{68})
\begin{equation} \label{69}
n_{\pm}(x) - n_{\pm} \sim {z_{\pm} \over 2x}
\end{equation}
still is of the form (\ref{67}), even when the
$n_{\pm}$ densities diverge.
Consequently, one can put
\begin{equation} \label{70}
n(x) - n = {2 z \cosh (\beta \varphi) \over (2 x)^{\beta/2}} f(2x) 
\end{equation}
where $f$ is a function regular in $\beta$ around $\beta = 2$,
with 
\begin{subequations} \label{71}
\begin{equation} \label{71a}
f(0) = 1
\end{equation}
The density $n(x)$ is supposed to tend to its asymptotical
$x\to\infty$ value $n$ faster than any inverse power of $x$,
so that
\begin{equation} \label{71b}
\lim_{x\to\infty} f(2x) \to 0 \quad {\rm faster\ than\ any\
inverse\ power\ of\ } x
\end{equation}
\end{subequations}
According to (\ref{45}), it holds
\begin{equation} \label{72}
-\beta z {\partial \gamma \over \partial z} = 
z \cosh (\beta \varphi) \int_0^{\infty} \rd t\, t^{-\beta/2}
f(t)
\end{equation}
An integration per partes gives
\begin{equation} \label{73}
\int_0^{\infty} \rd t\, t^{-\beta/2} f(t) = 
{1\over 1-\beta/2} \left[ t^{1-\beta/2} f(t)
\big\vert_{t=0}^{\infty} - \int_0^{\infty} \rd t\, t^{1-\beta/2}
{\partial f(t) \over \partial t} \right]
\end{equation}
For $\beta<2$, $t^{1-\beta/2}f(t)$ vanishes at $t=0$ 
as well as in the limit $t\to\infty$ due to the fast decay of
$f(t)$.
Then,
\begin{equation} \label{74}
\beta {\partial \gamma \over \partial z} = {\cosh(\beta \varphi)
\over 1-\beta/2 } \int_0^{\infty} \rd t\, t^{1-\beta/2} 
{\partial f(t) \over \partial t} 
\end{equation}
When $\beta\to 2^-$, one can perform a $(2-\beta)$ expansion 
of the integral in (\ref{74}),
\begin{equation} \label{75}
\int_0^{\infty} \rd t\, t^{1-\beta/2} {\partial f(t) 
\over \partial t} = f(\infty) - f(0)
+ O(2-\beta) 
\end{equation}
With regard to (\ref{71a}) and (\ref{71b}), 
one arrives at the desired formula (\ref{44}).

\section{Conclusion}
A two-dimensional model for the interface between an electrolyte and an
electrode has been considered: The 2D TCP bounded by a rectilinear ideal
conductor wall. 
Previously, the surface tension $\gamma$ in this model was known 
\cite{Forrester} only at the special inverse temperature $\beta =2$ (in
which case, for obtaining a finite result, a hard core repulsion between
the particles and the wall had to be assumed). 
Now, the main result of the present paper, eq. (\ref{41}), 
provides the surface tension for point particles at any temperature, 
in the stability range of the model $\beta <2$.

The surface tension depends on the bulk density $n$ as $n^{1/2}$, as a
priori expected for dimensional reasons. Its temperature dependence
is more complicated than the one of the bulk pressure but simpler than
the temperature dependence of some bulk thermodynamic quantities 
derived in I. 

\renewcommand{\theequation}{A\arabic{equation}}
\setcounter{equation}{0}

\section*{Appendix}
In this Appendix, the correction $K^{(1)}$ to the renormalized bond is
studied. 
In terms of the Green function $K^{(0)}$ which obeys 
\begin{equation} \label{A1}
(\Delta_1-\kappa^2)K^{(0)}(1,2) = 2 \pi \beta \delta(1,2)
\end{equation}
with the boundary condition $K^{(0)}(1,2)=0$ when point 1 is at
$x=0$, the solution of (\ref{61}) with the same boundary condition is
\begin{equation} \label{A2}
K^{(1)}(1,2) = \int \rd 3\,K^{(0)}(1,3)K^{(0)}(3,2)\delta n(3) 
\end{equation}
where $K^{(0)}$ is given by (\ref{52}) and $\delta n(3)=(1/2)n\beta
K_0(2\kappa x_3)$. 
Thus, the desired integral (\ref{62}) is  
\begin{equation} \label{A3}
\int_0^{\infty}\rd x_1K^{(1)}(1,1)=
\int_0^{\infty}\rd x_1\int_{-\infty}^{\infty}\rd y_3\int_0^{\infty}
\rd x_3 [-\beta K_0(\kappa r_{13})+\beta K_0(\kappa r_{13}^*)]^2
\frac{\beta n}{2}K_0(2\kappa x_3) 
\end{equation}

The integrals on $x_1$ and $y_3$ will be performed first. Since the
integrand is an even function of $x_1$, the integral on $x_1$ and 
$y_3$ is half that integral performed on the whole plane. 
Furthermore, $y_3$ can be shifted into $y_1$. 
Therefore,
\begin{equation} \label{A4}
I(x_3)=\int_0^{\infty}\rd x_1\int_{-\infty}^{\infty}\rd y_3 
[-K_0(\kappa r_{13})+K_0(\kappa r_{13}^*)]^2=
\int \rd^2 r_1[K_0^2(\kappa r_{13})-
K_0(\kappa r_{31})K_0(\kappa r_{13}^*)]
\end{equation}
One now has a convolution integral which can be performed by going
to Fourier space:
\begin{eqnarray} \label{A5}
I(x_3) & = & \int \rd^2 k\frac{1-J_0(2kx_3)}{(\kappa^2+k^2)^2} =
\frac{\pi}{\kappa^2}\left[1+\kappa\frac{d}{d\kappa}\int_0^{\infty}
\frac{\rd k\,kJ_0(2kx_3)}{\kappa^2+k^2}\right] \nonumber \\
& = & \frac{\pi}{\kappa^2}\left[ 1+\kappa\frac{dK_0(2\kappa x_3)}
{d\kappa} \right] = \frac{\pi}{\kappa^2}
\left[ 1-2\kappa x_3 K_1(2\kappa x_3) \right] 
\end{eqnarray}
Using (\ref{A5}) in (\ref{A3}) gives, using as a rescaled integration
variable $x=2\kappa x_3$, 
\begin{equation} \label{A6}
\int_0^{\infty}\rd x_1 K^{(1)}(1,1)=\frac{\beta^2}{8\kappa}\int_0^{\infty}
\rd x [1-xK_1(x)] K_0(x) 
\end{equation}
After an integration by parts using $K_1(x)=-dK_0(x)/dx$, and
taking into account (\ref{58b}), one finds (\ref{62}).

\section*{Acknowledgments}
The visit of Ladislav {\v S}amaj to Orsay is supported by NATO.
A partial support by Grant VEGA 2/7174/20 is acknowledged.

\end{document}